\documentclass[onecolumn,authoryear]{els-mrw} 

\usepackage{amsmath,amssymb,amsfonts,amsthm,makeidx,graphicx}
\usepackage{txfonts}
\usepackage{helvet}

\newcommand{\beq}{\begin{eqnarray}}
\newcommand{\eeq}{\end{eqnarray}}
\newcommand{\Slash}[1]{{\ooalign{\hfil/\hfil\crcr$#1$}}}

\newcommand{\nn}{\nonumber \\}

\begin{document}

\chapter{Spin Structure of the Nucleon: Overview}\label{chap1}

\author[1,2]{Yoshitaka Hatta}

\address[1]{\orgname{Brookhaven National Laboratory}, \orgdiv{Physics Department}, \orgaddress{Upton, NY 11973, USA}}
\address[2]{\orgname{Brookhaven National Laboratory}, \orgdiv{RIKEN BNL Research Center}, \orgaddress{Upton, NY 11973, USA}}


\maketitle




\begin{abstract}[Abstract]
I present a pedagogical review of the decomposition of the proton spin. Both the  Jaffe-Manohar and Ji decompositions are discussed. Particular emphasis is placed on the quark and gluon orbital angular momenta, including their gauge invariant definitions, small-$x$ behavior and connection to experimental observables.  
\end{abstract}

\begin{abstract}[Keywords]
 Deep Inelastic Scattering, Electron-Ion Collider, energy momentum tensor,  helicity,  gauge invariant spin decomposition,  Jaffe-Manohar decomposition, Ji decomposition,  orbital angular momentum, parton, proton spin, Quantum Chromodynamics, quark and gluon,   spin at small-$x$
\end{abstract}

\begin{glossary}[Key points]
\begin{itemize}
\item
Formulate the partonic decomposition of the proton spin 

\item
Describe the gauge invariant completion of the Jaffe-Manohar decomposition

\item
Review the derivation of the Ji decompotion and its connection to the generalized parton distribution

\item
Review the modern understanding of  parton orbital angular momentum and challenges for future
\end{itemize}
\end{glossary}

\begin{glossary}[Nomenclature]
\begin{tabular}{@{}lp{34pc}@{}}
DIS & Deep Inelastic Scattering\\
OAM &Orbital Angular Momentum\\
EMC & European Muon Collaboration\\
EIC & Electron-Ion Collider\\
PDF &Parton Distribution Function\\
GPD & Generalized Parton Distribution
\end{tabular}
\end{glossary}

\section{Introduction}

The proton was discovered by Rutherford  as a fundamental constituent of atomic nuclei in 1919 when the concept of spin was nonexistent. During the 1930s, the proton was established as a spin-$\frac{1}{2}$ particle (in units of the Planck constant $\hbar=\frac{h}{2\pi}$) through the measurement of its magnetic moment. Quickly thereafter, proton (and neutron) spin rapidly became a cornerstone of nuclear  physics and beyond. It underlies the hyperfine structure of atomic energy levels, determines nuclear magnetic resonance, and governs the behavior of matter in external electromagnetic fields. At the nuclear level, proton spin plays a central role in determining nuclear structure, selection rules, and reaction dynamics.  

Even in these early days of quantum mechanism, one could already ask a fundamental question: Where does the proton spin $\frac{1}{2}$ come from in the first place? This was a legitimate question because the proton was already known to be a composite particle  at that time, as  evident from the measured value of the proton magnetic moment, which is 2.79 times larger than what is  expected for  an elementary Dirac particle. According to quantum mechanics, the total angular momentum of a composite system arises from the sum of spins and orbital angular momenta of its constituents. However, without knowing what these fundamental degrees of freedom are, it was not possible to precisely formulate  the question.    

With the advent and establishment of Quantum ChromoDynamics (QCD) in the 1960s-1970s, it has been understood that the proton is a composite, relativistic bound state of quarks and gluons, the fundamental particles  of the strong interaction.  Understanding how its spin emerges from the dynamics and angular momentum of these constituents has become one of the most enduring and challenging problems in hadronic physics. 
This conceptual challenge moved to the forefront of high-energy physics in the late 1980s, when deep-inelastic scattering (DIS) experiments began to probe the spin structure of the proton at the partonic level. In 1988, the European Muon Collaboration (EMC) reported measurements of the polarized structure function 
$g_1(x)$ indicating that the contribution of quark helicities to the proton spin is unexpectedly small. Interpreted within the parton model, the EMC data suggested that the sum of quark and antiquark spins accounts for only a minor fraction of the proton’s total spin, in striking contrast with naive constituent quark models in which the proton spin is predominantly carried by valence quarks. This result, which became known as the “proton spin crisis,” fundamentally altered the field. It revealed that relativistic effects, gluon polarization, and orbital angular momentum (OAM)  must play an essential role, and it exposed the limitations of simple pictures of nucleon structure. The EMC discovery triggered  an extensive experimental and theoretical program that continues to this day, encompassing polarized lepton–nucleon scattering, proton–proton collisions, lattice QCD, and the development of gauge-invariant spin decomposition frameworks within QCD.

In this overview, I will present the modern understanding of the   proton spin decomposition problem based on QCD. There are already many reviews on this subject, e.g.,      
\cite{Cheng:1996jr,Kuhn:2008sy,Leader:2013jra,Ji:2020ena}. I will mostly focus on  theoretical developments over the past decade, leaving phenomenology, including the physics of transverse polarization, to other chapters of this volume. 

\section{Spin crisis}\label{chap1:sec1}

Imagine a proton moving in the positive $z$ direction and its spin is pointing in the same direction. Such a proton is called `longitudinally polarized,' and has helicity $+\frac{1}{2}$ (in units of $\hbar$) defined as the projection of spin along the direction of motion.    Since the proton is a bound state of quarks, antiquarks and gluons, one should be able to decompose the value $\frac{1}{2}$ into four contributions 
\beq
\frac{1}{2}=\frac{1}{2}\Delta\Sigma + \Delta G + L^q+L^g. \label{1}
\eeq
$\frac{1}{2}\Delta\Sigma$ is the total helicity contribution of quark and antiquarks, and $\Delta G$ is that of gluons.  
$L^{q}$ and $L^g$ are the $z$-components of the orbital angular momenta (OAM) of quarks and gluons, respectively. 
The decomposition (\ref{1}) can be done rigorously in QCD as I will discuss later.  
The earliest and simplest insight into how the decomposition works in practice came from the quark model (\cite{Chodos:1974pn}).  In this model, the proton consists of two up quarks and one down quark. Gluons are entirely absent, so that  $\Delta G=L^g=0$. In the nonrelativistic version of the model, the proton single-particle state is given by  
\beq
|P,+\rangle = \frac{1}{3\sqrt{2}}\Bigl[\{2|++-\rangle +|+-+\rangle -|-++\rangle \} + {\rm cyclic\ permutations} \Bigr],
\eeq
where $\pm$ refers to the helicity $\pm \frac{1}{2}$ of the three  constituent  quarks. One immediately concludes that $m_z=\frac{1}{2}$ is decomposed as $\frac{1}{2}=\frac{1}{2}+\frac{1}{2}-\frac{1}{2}$, namely, two quarks are spin-up and one quark is spin-down. This means that $\Delta\Sigma=1$, or the proton spin comes 100\% from the quark helicity. A more sophisticated treatment includes  relativistic corrections.  Assume that the quarks are moving inside the proton under the influence of some centrifugal  potential. The spherically symmetric, ground state solution of the Dirac equation takes the form 
\beq
\psi=\frac{1}{\sqrt{4\pi}}\begin{pmatrix} f(r)\chi \\ -ig(r)\hat{r}\cdot \vec{\sigma} \chi
\end{pmatrix}
\eeq
where $f,g$ are normalized as  $1=\int_0^\infty dr r^2(f^2(r)+g^2(r))$. $\chi=\begin{pmatrix} 1 \\ 0 \end{pmatrix}$ is the two-component spinor representing the $m_z=+\frac{1}{2}$ state and $\vec{\sigma}=(\sigma^1,\sigma^2,\sigma^3)$ are the Pauli matrices. The upper component has OAM  $l=0$ and the lower component has $l=1$. In this setup, $\Delta \Sigma$ is given by the  expectation value 
\beq
\Delta\Sigma=\int d^3r \psi^\dagger \begin{pmatrix}\sigma^z & 0 \\ 0 & \sigma^z\end{pmatrix} \psi 
=1 -\frac{4}{3}\int dr r^2g(r) .\label{rela}
\eeq
This shows that, due to relativistic effects, part of helicity is converted to OAM (\cite{Sehgal:1974rz}). Simple model calculations such as  the bag model  suggest that $\Delta\Sigma\sim 0.7$. Namely, 70\% of the proton spin comes from quark spins, and the rest (30\%) is the OAM  $L^q$ of quarks.  
One might think that this is an oversimplified picture. However, empirically it was known that the quark model provided decent predictions for various properties of hadrons such as the mass spectrum and magnetic moment. Therefore, it was anticipated  that the true value of $\Delta \Sigma$ would not be very far off from the naive estimate  $\Delta\Sigma = 0.7 \sim 1$. 

However, this expectation was completely shattered in 1987 when the European Muon Collaboration (EMC) at CERN reported  the first experimental extraction of $\Delta \Sigma$ from polarized Deep Inelastic Scattering (DIS)  (\cite{EuropeanMuon:1987isl}). In this process, the incoming muon emits a virtual photon with momentum $q^\mu$ and virtuality $Q^2=-q^2>0$ which then interacts with the proton with momentum $P^\mu$. They have measured double spin asymmetry
\beq
A_{LL} = \frac{d\sigma_{++}-d\sigma_{+-}}{d\sigma_{++}+d\sigma_{+-}},
\eeq
where $\sigma_{ss'}$ denotes the inclusive cross section with proton helicity $s$ and the muon helicity $s'$. From the ratio, one can extract the $g_1(x)$ structure function as the function of the Bjorken variable $x=\frac{Q^2}{2P\cdot q}$ (at the renormalization scale $Q^2\approx 10.7$ GeV$^2$). The integral of $g_1(x)$ is related to $\Delta \Sigma$ as 
\beq
\int_0^1 dx g_1(x) = \frac{1}{9}\Delta \Sigma + \frac{1}{12}(\Delta u-\Delta d)+\frac{1}{36}(\Delta u + \Delta d -2\Delta s)+{\cal O}(\alpha_s)
\eeq
Here, $\Delta f$ ($f=u,d,s$) is the difference between the numbers of spin-up and spin-down quarks and antiquarks with flavor $f$.  $\frac{1}{2}\Delta\Sigma=\frac{1}{2}(\Delta u + \Delta d +\Delta s)$ then represents  the total net polarization of all quarks and antiquarks. The isovector combination $\Delta u-\Delta d$ is equal to the well-known quantity $g_A^{(3)}$, the isovector axial coupling of the nucleon. Thanks  to the flavor SU(3) symmetry, the octet combination $\Delta u+\Delta d-2\Delta s$ happens to be related to the amplitude of  the hyperon semileptonic decay $\Xi^-\to \Lambda + e^-+\bar{\nu}_e$ which can be fixed  experimentally. Subtracting these and neglecting the ${\cal O}(\alpha_s)$ next-to-leading order (NLO) corrections, the EMC collaboration concluded that 
\beq
\Delta \Sigma = 0.12\pm 0.09\pm 0.14 .\label{emc}
\eeq
Namely, the quark contribution is significantly less than unity, only 10\%, even consistent with zero considering the error bars. This was in  striking disagreement with the quark model prediction $\Delta \Sigma \sim 1$ and sent a shockwave to the hadron physics community. The phenomenon was dubbed the `spin crisis' and effectively marked the birth of a new field,  QCD spin physics. 
In fact, the initial result (\ref{emc}) was an underestimation. With better measurements of the small-$x$ region and inclusion of perturbative QCD corrections,  the currently accepted value is larger $\Delta \Sigma \sim 30$\%. This is however  still significantly below unity, and the puzzle continues to date.

As a matter of course, determining the gluon spin contribution $\Delta G$ has been a central goal of spin physics in the decades that followed. In practice, $\Delta G$ is obtained by integrating the polarized (helicity) gluon parton distribution function (PDF) $\Delta G(x)$ 
\beq
\Delta G=\int_0^1 dx \Delta G(x),
\eeq
which in turn can be determined via a global analysis. 
Significant progress has been made at the Relativistic Heavy Ion Collider (RHIC) at Brookhaven National Laboratory in the U.S.  through polarized proton–proton collisions (\cite{Bunce:2000uv}). The PHENIX and STAR collaborations measured double spin asymmetries in jet, hadron and direct photon productions, which are sensitive to the gluon helicity PDF $\Delta G(x)$ over a specific range of $x$ 
\cite{PHENIX:2008swq,STAR:2021mfd,PHENIX:2022lgn}. 
Current global fits incorporating RHIC data indicate a positive gluon contribution in the measured 
$x$-range $x>0.05$  (\cite{deFlorian:2024utd,Hunt-Smith:2024khs}), with a magnitude comparable to or possibly exceeding the quark contribution in that region. However, the total integral  
$\int _0^1 dx \Delta G(x)$ remains uncertain due to the poorly constrained small-$x$ region. This unresolved domain could significantly modify the total gluon contribution. 

One of the major goals of the Electron-Ion Collider (EIC,  \cite{AbdulKhalek:2021gbh}) to be constructed at Brookhaven National Laboratory is to  understand the origin of the proton spin, which roughly amounts to  determining all four terms in (\ref{1}). In particular, the current large uncertainty in $\Delta G$ is expected to be significantly reduced through a precise measurement of the $g_1(x)$ structure function down to $x\sim  10^{-4}$. The EIC will also be the first machine to access the OAM  contributions $L^{q,g}$ in (\ref{1}). This will be briefly discussed toward the end of this review. Before doing so, first I need to explain the theoretical foundation and definition  of OAM in QCD.

\section{Jaffe-Manohar decomposition }

In quantum field theory, the description of the spin and angular momentum of an elementary particle or a bound state is based on the angular momentum tensor $M^{\mu\nu\lambda}$ which is the Noether current associated with  Lorentz symmetry $x^\nu\to x^\nu+\omega^{\nu\lambda}x_\lambda$. It can be derived from the QCD Lagrangian 
\beq
{\cal L} = \bar{\psi}(i\Slash D-m)\psi -\frac{1}{4}F_{\mu\nu}F^{\mu\nu},
\eeq
via the standard method and reads 
\beq
M_{can}^{\mu\nu\lambda}=x^\nu T_{can}^{\mu\lambda}-x^\lambda T_{can}^{\mu\nu} -\frac{1}{2}\underbrace{\epsilon^{\mu\nu\lambda\rho}\bar{\psi}\gamma_\rho \gamma_5\psi}_{\rm quark \ helicity} + \underbrace{F^{\mu\lambda}A^\nu -F^{\mu\nu}A^\lambda}_{\rm gluon\  helicity} \label{m}
\eeq
where $T^{\mu\rho}_{can}$ is the  energy momentum tensor 
\beq
T^{\mu\nu}_{can}=\underbrace{\bar{\psi}\gamma^\mu i\overleftrightarrow{\partial}^\nu \psi}_{\to {\rm quark\  OAM}} -\underbrace{F^{\mu\alpha}\partial^\nu A_\alpha}_{\to {\rm gluon \ OAM}} -\eta^{\mu\nu}{\cal L} \label{t}
\eeq
with $\eta^{\mu\nu}={\rm diag}(1,-1,-1,-1)$, $\psi=(u,d,s,.\cdots)^T$ and $\overleftrightarrow{\partial}^\mu = \frac{1}{2}(\partial^\mu - \overleftarrow{\partial}^\mu)$. My convention of the gamma matrices are $\gamma^\mu=\begin{pmatrix} 0 & \sigma^\mu \\ -\bar{\sigma}^\mu & 0 \end{pmatrix}$ and $\gamma_5=\begin{pmatrix} -1 & 0 \\ 0 & 1\end{pmatrix}$. The subscript ${\it can}$ means that these are the canonical energy/angular momentum tensors. In (\ref{m}) and (\ref{t}), one can immediately recognize four structures,   in one-to-one correspondence with the four terms in (\ref{1}). The first term in $T^{\mu\nu}_{can}$ represents the quark momentum. Its cross product with $x^\nu$ as prescribed in $M_{can}^{\mu\nu\lambda}$ makes it the quark OAM $L^q$. Similarly, the second term in $T_{can}^{\mu\nu}$ represents the gluon momentum, constituting the gluon OAM $L^g$.  The explicit connection is established  by taking the forward proton matrix element of the operators $\langle PS|...|PS\rangle$ in the proton single particle state characterized by the momentum 4-vector $P^\mu$ and the spin 4-vector 
\beq
2S^\mu = \bar{u}(PS)\gamma^\mu\gamma_5 u(PS),
\eeq
where $u$ is the proton's Dirac spinor. For a proton moving in the $z=x^3$ direction, one takes the  $\mu\nu\lambda=+12$ component\footnote{ The light-cone coordinates are defined as  
\beq
x^\pm =\frac{1}{\sqrt{2}}(x^0\pm x^3), \qquad \gamma^\pm = \frac{1}{\sqrt{2}}(\gamma^0\pm \gamma^3) \qquad g^{+-}=g_{-+}=1, \qquad \epsilon^{-+12}=\epsilon^{12}=1
\eeq
} and find the correspondence 
\beq
&&2\Delta \Sigma S^+ =\langle PS|\bar{\psi}\gamma^+\gamma_5\psi|PS\rangle, \nn 
&&2\Delta GS^+ =  \langle PS|\epsilon^{ij}F^{i+}A^j|PS\rangle, \nn 
&&2L_{can}^qS^+= \langle PS|\epsilon^{ij}x^i \bar{\psi} \gamma^+i\overleftrightarrow{\partial}^j \psi|PS\rangle, \nn 
&& 2L_{can}^gS^+= \langle PS|\epsilon^{ij}x^i F^{\alpha+}\partial^j A_\alpha|PS\rangle,
\label{jm} 
\eeq
where $i,j=1,2$ denote the coordinates transverse to the proton direction of motion ($+x^3$ direction). Eq.~(\ref{jm}) is called the Jaffe-Manohar decomposition (\cite{Jaffe:1989jz}). 

While intuitively appealing, the Jaffe-Manohar decomposition has an obvious deficiency. Except for $\Delta\Sigma$, the operators involved are not gauge invariant. One has to fix a gauge when evaluating $\Delta G$ and  $L_{can}^{q,g}$, or else they vanish due to Elitzur's theorem. In  $L_{can}^q$, it is tempting to replace the ordinary derivative with the covariant derivative $\partial^i\to D^i$ to make it gauge invariant. But this ad hoc prescription does not make  $L_{can}^g$ gauge invariant. Then which gauge should one choose? Originally, Jaffe and Manohar argued that  the operators should be  evaluated in the light cone gauge 
\beq
A^+=0.
\eeq
Indeed, in the light-cone gauge, QCD looks closest to Feynman's parton picture. The physical interpretation of the `partonic' decomposition of the proton spin is most transparent in this gauge. However, from field theory perspective, all observables must have a gauge invariant definition. In fact, the gauge invariant generalization of $\Delta G$ has been known for a long time   (\cite{Manohar:1990jx}), together with the gauge invariant  gluon helicity PDF  
\beq
\Delta G = \int_0^1 dx \Delta G(x), \qquad \Delta G(x) =  \frac{i}{xS^+}\int \frac{dy^-}{2\pi} e^{ixP^+y^-}\langle PS|F^{+\alpha}(0)U_{0y}\tilde{F}^+_{\ \alpha}(y^-)|PS\rangle ,\label{g}
\eeq
where $\tilde{F}^{\mu\nu}=\frac{1}{2}\epsilon^{\mu\nu\alpha\beta}F_{\alpha\beta}$.  A Wilson line  along the light-cone $U_{0y}$ is necessary to make the nonlocal operator gauge invariant.  It is easy to see that, if one takes the light-cone gauge where $F^{+\alpha}=\partial^+A^\alpha$, the local operator in (\ref{jm}) is recovered. 
However, in generic gauges, the operator is nonlocal in the light-cone direction, even after integrating over $x$
\beq
\int_0^1 dx \Delta G(x) = \frac{1}{2S^+}\int dy^-\theta(y^-)\langle PS|F^{+\alpha}(0)U_{0y}\tilde{F}^+_{\ \alpha}(y^-)|PS\rangle, \label{intg}
\eeq
where I used the prescription   $\frac{1}{x}\to \frac{1}{x-i\epsilon}$  to perform the $x$-integral. (The result does not depend on the choice of  the $i\epsilon$-prescription.) 

The issue of gauge invariance for the OAM $L_{can}^q$ and $L_{can}^g$,  as well as spin decomposition in general, has been a subject of longstanding debate. An early discussion in \cite{Bashinsky:1998if} actually  came close the final form presented below. Meanwhile, a proposal by  \cite{Chen:2008ag} triggered a substantial controversy in the QCD spin physics community, see e.g.,  \cite{Wakamatsu:2013voa,Leader:2013jra} and references therein. They  argued that the gauge field consists of the `physical' and `pure gauge' parts 
\beq
A^\mu = A^\mu_{phys} + A^\mu_{pure}.
\eeq
Under gauge transformations, the physical part transforms homogeneously, while the pure gauge part transforms inhomogeneously 
\beq
A^\mu_{phys} \to UA^\mu_{phys}U^\dagger  \qquad A^\mu_{pure} \to UA^\mu_{pure} U^\dagger-\frac{i}{g} U\partial^\mu U^\dagger. \label{inh}
\eeq
Accordingly, define 
\beq
D^\mu_{pure} = \partial^\mu +igA_{pure}^\mu = D^\mu -igA^\mu_{phys}.
\eeq
It is then easy to check that the following version of the Jaffe-Manohar decomposition  
\beq
&&\Delta G\sim \langle PS|\epsilon^{ij}F^{i+}A^j_{phys}|PS\rangle, \nn 
&&L_{can}^q\sim \langle PS|\epsilon^{ij}x^i \bar{\psi} \gamma^+iD_{pure}^j \psi|PS\rangle, \label{chen}  \\
&& L_{can}^g\sim \langle PS|\epsilon^{ij}x^i F^{\alpha+}D_{pure}^j A^{phys}_\alpha|PS\rangle, \notag
\eeq
is gauge invariant.  
This seems like an elegant construction, but an obvious question is: What are $A^\mu_{phys}$ and $A^\mu_{pure}$ exactly? \cite{Chen:2008ag} postulated that they should satisfy the following conditions 
\beq
 \vec{D}_{pure}\times \vec{A}_{pure}=0, \qquad [\vec{A}_{phys}, \vec{E}]=0,
\eeq
with $E^i=F^{i0}$. This is analogous to the Coulomb gauge condition $\vec{\nabla}\cdot \vec{A}=0$ in QED. 
However, with this choice, $\Delta G$ defined by (\ref{chen}) does not coincides with the $\Delta G$ that has been measured in high energy experiments. Instead, 
\cite{Hatta:2011zs} chose 
\beq
A^\mu_{phys}(y^-,\vec{y}) = -\int dw^-\theta(w^--y^-) U_{yw}F^{+\mu}(w^-,\vec{y}),  \label{hat}
\eeq
which manifestly satisfies (\ref{inh}).  (\ref{hat}) is chosen such that, when substituted in (\ref{chen}), the resulting $\Delta G$ coincides with the standard definition (\ref{intg}) in QCD spin physics. Of course, this is just a reinterpretation of the known formula for $\Delta G$. But the benefit  of the new  formulation is that, once $A_{phys}$ is fixed in this way,  it uniquely predicts the gauge invariant generalization of the OAMs $L_{can}^{q}$ and $L_{can}^g$ by simply plugging the same $A^\mu_{phys}$ and the corresponding $D_{pure}$ into (\ref{chen}).  (\ref{chen}) together with  (\ref{hat}) achieves the gauge invariant completion of the Jaffe-Manohar decomposition (\ref{1}) relevant to high energy QCD spin physics.  In the light-cone gauge $A^+=0$, it reduces to (\ref{jm}).

\section{Ji decomposition} 

The original Jaffe-Manohar decomposition is not gauge invariant because already  the canonical  energy momentum tensor $T^{\mu\nu}_{can}$ 
(\ref{t}) is not gauge invariant. Moreover, the tensor is not symmetric in the two indices $\mu,\nu$, and is conserved only with respect to the first index, $\partial_\mu T^{\mu\nu}_{can}=0$, not the second $\partial_\nu T^{\mu\nu}_{can}\neq 0$.  However, there is a standard method to `improve' the energy momentum tensor. Let us write (\ref{m}) as 
\beq
M^{\mu\nu\lambda}=x^\nu T^{\mu\lambda}-x^\lambda T^{\mu\nu} + H^{\mu\nu\lambda},
\eeq
Define 
\beq
G^{\rho\mu\nu} \equiv  \frac{1}{2}(H^{\rho\mu\nu}-H^{\mu\rho\nu}-H^{\nu\rho\mu}), 
\eeq
and 
\beq
T^{\mu\nu}_{Bel} \equiv T^{\mu\nu}_{can} + \partial_\rho G^{\rho\mu\nu}
\eeq
This is known as the Belinfante-improved energy momentum tensor. One can check that it is symmetric in $\mu,\nu$ and conserved with respect to both indices. The added term is a total derivative, so it does not affect the total energy (hadron masses). Using the QCD equations of motion, one can show that 
\beq
T_{Bel}^{\mu\nu}= \bar{\psi}\gamma^{(\mu}i\overleftrightarrow{D}^{\nu)}\psi -F^{\mu\rho}F^\nu_{\ \rho}+\frac{g^{\mu\nu}}{4}F^{\alpha\beta}F_{\alpha\beta} \label{belt}
\eeq
where $A^{(\mu}B^{\nu)}\equiv \frac{1}{2}(A^\mu B^\nu + A^\nu B^\mu)$. 
Similarly, the angular momentum tensor can be   improved 
\beq
M^{\mu\nu\lambda}_{Bel}= x^\nu T^{\mu\lambda}_{Bel}-x^\lambda T^{\mu\nu}_{Bel} \label{belm}
\eeq
Alternatively, $T^{\mu\nu}_{Bel}$ can be derived in a general relativity way, by first embedding the QCD action in a curved space and then functionally differentiating it with respect to the metric
\beq
T_{\mu\nu}=\frac{2}{\sqrt{-g}}\frac{\delta}{\delta g^{\mu\nu}} \int d^4x \sqrt{-g} {\cal L}\,[g^{\mu\nu},\psi,A^\mu]. 
\eeq
In practice, this second method is much simpler. 

In (\ref{belt}), one can clearly identify the quark and gluon parts of the energy momentum tensor that are separately gauge invariant.  (\ref{belm}) then achieves the  decomposition of the angular momentum into the quark and gluon parts in terms of local, gauge invariant operators. A further decomposition is possible in the quark part by using the identity 
\beq
\bar{\psi}\gamma^{(\mu}i\overleftrightarrow{D}^{\nu)}\psi = \bar{\psi}\gamma^{\mu}i\overleftrightarrow{D}^{\nu}\psi -\frac{1}{4}\epsilon^{\mu\nu\rho\sigma}\partial_\rho (\bar{\psi}\gamma_\sigma \gamma_5\psi).
\eeq
After taking the cross product with $x^\lambda$, the first term can be interpreted as the quark orbital angular momentum featuring the covariant derivative $D^\nu$. This is in contrast to  the Jaffe-Manohar decomposition which features the ordinary derivative $\partial^\nu$ or its gauge covariant generalization $D^\nu_{pure}$. 
In the second term, one can integrate by parts  $x^\lambda \partial_\rho (\bar{\psi}\gamma_\sigma \gamma_5\psi) \to -\delta^\lambda_\rho\bar{\psi}\gamma_\sigma \gamma_5\psi$ and recover  the familiar quark spin operator. 
One thus arrives at the  Ji sum rule (\cite{Ji:1996ek})
\beq
\frac{1}{2}= \sum_q J_q+J_g =\frac{1}{2}\Delta \Sigma + \sum_q L_{kin}^q + J_g .
\eeq
The subscript `{\it kin}' denotes the `kinetic' OAM, in distinction from the canonical OAM $L_{can}^q$ discussed earlier. 
Compared to the Jaffe-Manohar sum rule, only the quark helicity $\Delta\Sigma$ is common. 
The gluon angular momentum $J_g$ cannot be further decomposed into the helicity and orbital angular momentum part in terms of local gauge invariant operators. 

Let us take a loser look at the relevant matrix elements. Since the Ji sum rule is not tied to light-cone physics, one can simply work in the rest frame of the proton. Taking the $\mu\nu\lambda=012$ component of the angular momentum tensor and sandwiching between the proton states, one can write  
\beq
\langle PS|J_{q,g}^3|PS\rangle = \frac{1}{V}\langle PS|\epsilon^{ij}\int d^3x x^i T^{0j}_{q,g}(x)|PS\rangle 
=-i\lim_{\Delta \to 0}\epsilon^{ij}\frac{\partial}{\partial \Delta^i}\langle P'S|T^{0j}_{q,g}(0)|PS\rangle , \label{eps}
\eeq
where $V=\int d^3x$ is the volume integral. In order to get rid of the factor of $x^i$, a useful trick is to consider a slightly off-forward matrix element with $\Delta^\mu = P'^\mu-P^\mu$. Then $\langle T^{0i}(x)\rangle =\langle e^{i\hat{P}\cdot x}T^{0i}(0)e^{-\hat{P}\cdot x}\rangle= e^{i\Delta \cdot x}\langle T^{0i}(0)\rangle$. This allows one to  trade $x^i$ with the derivative with respect to $\Delta^i$. The off-forward matrix elements of the quark and gluon parts of the energy momentum tensor are parametrized as (\cite{Ji:1996ek}) 
\beq
\langle P'S'|T_{q,g}^{\alpha\beta}|PS\rangle &=& \bar{u}(P'S')\left[A_{q,g}(t)\gamma^{(\alpha}\bar{P}^{\beta)}+B_{q,g}(t)\frac{\bar{P}^{(\alpha}i\sigma^{\beta)\lambda}\Delta_\lambda}{2m_N} + D_{q,g}(t)\frac{\Delta^\alpha \Delta^\beta-\eta^{\alpha\beta}\Delta^2}{4m_N} + \bar{C}_{q,g} (t)m_N \eta^{\alpha\beta}\right]u(PS)  
\eeq
where $t=\Delta^2$ and $\bar{P}=\frac{P+P'}{2}$. 
$A,B,D,\bar{C}$ are called the gravitational form factors, defined separately for the quark and gluon parts. (The notation $C_{q,g}=D_{q,g}/4$ was used in the original paper \cite{Ji:1996ek}.) In analogy to the electromagnetic form factors that describe the interaction between the hadron and external electromagnetic fields, the gravitational form factors describe the coupling of the hadron to external gravitational fields, although in practice the latter interaction is too weak to be directly measurable. The $\bar{C}$ form factor is there because the quark and gluon parts are not separately conserved. Since the sum is conserved $\partial_\mu(\sum_q T^{\mu\nu}_q+T^{\mu\nu}_g)=0$, one has $\sum_q \bar{C}_q(t)+\bar{C}_g(t)=0$ for any $t$. The derivative in (\ref{eps}) acts on the manifest $\Delta$ factor in the $B$-term. Although not manifest, the spinor bilinear that multiplies the $A$-form factor also contains a term linearly dependent on $\Delta$. One thus obtains an alternative form of the Ji sum rule 
\beq
J_q=\frac{1}{2}(A_q(0)+B_q(0)), \qquad J_g=\frac{1}{2}(A_g(0)+B_g(0)) .\label{jisum}
\eeq
$A_i(0)$ has the physical meaning as the momentum fraction of the proton carried by the parton species $i$. As such, they satisfy the sum rule $\sum_q A_q(0)+A_g(0)=1$. This immediately implies that $\sum_q B_q(0)+B_g(0)=0$, which represents angular momentum conservation. $A$'s are nothing but the second momentum of the unpolarized quark and gluon PDFs 
\beq
A_q=\int_{-1}^1 dx x q(x) ,\qquad A_g=\int_0^1 dx x G(x),
\eeq
that have been  extracted from global analysis. Thus the main objects of interest are the values  $B_{q,g}(0)$ which are quite challenging to determine experimentally. At the moment, the only known method to constrain $B$'s in experiments is to exploit its connection to the generalized parton distributions (GPDs). GPDs are off-foward generalization of PDFs defined as 
\beq
&&\bar{P}^+\int \frac{dz^-}{2\pi}e^{ix\bar{P}^+z^-}\langle P'S'|\bar{q}(-z^-/2)\gamma^+U_{-z/2,z/2}q(z^-/2)|PS\rangle = \bar{u}(P'S')\left[\gamma^+H_q(x,\xi,t)+\frac{i\sigma^{+\nu}\Delta_\nu}{2m_N} E_q(x,\xi,t)\right]u(PS),
\label{gpd}
\eeq
and similarly for the gluon. The parameter $\xi=-\frac{\Delta^+}{2\bar{P}^+}$ is called skewness. 
Multiplying both sides by $x$ and integrating over $x$, on the left hand side, one gets 
\beq
\langle P'S'|T^{++}_q|PS\rangle = 
\bar{u}(P'S')\left[\bar{P}^+A_q(t)\gamma^+ +\bar{P}^+B_q(t)\frac{i\sigma^{+\lambda}\Delta_\lambda}{2m_N} + D_q(t) \frac{(\Delta^+)^2}{4m_N} \right]u(PS)
\eeq
The right hand side leads to the second moment of $H_q$ and $E_q$. This gives a yet another form of the Ji sum rule 
\beq
\int_{-1}^1 dx xH_q(x,\xi,t)= A_q(t)+\xi^2D_q(t), \qquad \int_{-1}^1 dx xE_q(x,\xi,t) = B_q(t)-\xi^2D_q(t),\label{jisum2}
\eeq
and similarly for the gluon GPDs. 
GPDs can be probed in  exclusive processes such as deeply virtual Compton scattering (DVCS) at facilities like Jefferson Laboratory and COMPASS at CERN.   
To determine $B_{q,g}(t)$, one must  extract the E-type GPDs $E_{q,g}(x,\xi,t)$ from these experiments, integrate over $x$ and extrapolate to $\xi,t\to 0$. Despite many years of effort, the extraction of $E_q$,  and especially $E_g$,  has proven to be quite challenging. This remains one of the major motivations for the future EIC experiment.   

\section{Orbital angular momentum and the Wigner distribution} 

As explained above, the OAM in the Jaffe-Manohar and Ji decompositions are different. The difference $L_{can}^q-L^q_{kin}$ is  sometimes referred to as the `potential' angular momentum (\cite{Wakamatsu:2010qj}). To gain  physical insight into this difference, let us introduce the Wigner distribution. 
Classically, OAM is the cross product of the position and momentum vectors $\vec{x}\times \vec{P}$. It is well known in quantum mechanics that  one can define a phase space distribution, called the Wigner distribution, which depends on both these variables $W(\vec{x},\vec{p})$,  despite an apparent conflict with the uncertainty principle. This idea can be straightforwardly generalized to hadron structure physics. The quark Wigner distribution reads  (\cite{Belitsky:2003nz})
\beq
W_q(x,b_\perp, k_\perp)= \int \frac{dy^- d^2y_\perp}{16\pi^3}\frac{d^2\Delta_\perp}{(2\pi)^2}e^{ixP^+y^- -ik_\perp \cdot y_\perp -i\Delta_\perp \cdot b_\perp}\langle P'|\bar{\psi}(-y/2)\gamma^+U_{-y/2,y/2}\psi(y/2)|P\rangle. \label{wilson}
\eeq
The gluon Wigner distribution can be defined similarly. 
This is a quasi-probabilistic distribution in the sense that it is not positive definite everywhere. Still, as in quantum mechanics, it serves as the `mother distribution' of lower dimensional distributions. By integrating over $k_\perp$, one recovers the GPD $H_q(x,b_\perp)$. By integrating over $b_\perp$, one recovers the transverse momentum dependent (TMD) distribution $f_q(x,k_\perp)$. 
It then  makes perfect sense to define OAM as (\cite{Lorce:2011kd,Hatta:2011ku,Lorce:2011ni})
\beq
L_q= \int dx \int d^2b_\perp d^2k_\perp (b_\perp \times k_\perp)W_q(x,b_\perp,k_\perp). \label{wigner}
\eeq
As long as the Wigner distribution is gauge invariant, (\ref{wigner}) is gauge invariant. But {\it which} OAM is this? The answer depends on the Wilson line $U$  unspecified in (\ref{wilson}). Let us first consider a staple-shaped Wilson line that extends to light-cone infinity
\beq
U^{staple}_{-y/2,y/2}=U_{-y/2, \infty}(-y_\perp/2)U_{ \infty}(-y_\perp/2,y_\perp/2)U_{ \infty y/2}(y_\perp/2)
\eeq
This results in the canonical orbital angular momentum 
\beq
\int d^2k_\perp b_\perp \times k_\perp W_q^{staple}(x,b_\perp,k_\perp) =\langle \bar{\psi}b_\perp \times (-i D^{pure}_\perp)\psi \rangle.
\eeq
To derive this, replace $k_\perp$ with a derivative acting on $y_\perp$ using the phase factor $e^{-ik_\perp \cdot y_\perp}$  and then use the formula 
\beq
\frac{\partial}{\partial y_\perp^i} U_{\infty y^-}[y_\perp]\psi(y) =U_{ \infty, y^-}[y_\perp] D_{\perp i}^{pure}\psi(y)
\eeq
On the other hand, if one chooses a straight Wilson line that directly connects the two points $\frac{y}{2}$ and $-\frac{y}{2}$, one finds the kinetic OAM (\cite{Ji:2012sj}). Clearly, the difference comes from a closed Wilson loop that extends to the light-cone infinity $y^-=\infty$, resulting in the operator (cf. (\ref{hat}))  
\beq
\epsilon^{ij} b^i_\perp \int dy^- F^{+j} .\label{tor}
\eeq
In terms of the color electric and magnetic fields, one can write 
\beq
F^{+j} = \frac{1}{\sqrt{2}}(\vec{E}+\vec{v}\times \vec{B})^j
\eeq
with velocity $\vec{v}=(0,0,-1)$. This is nothing but  the color Lorentz force. Therefore,   
(\ref{tor}) can be interpreted as torque that a quark experiences as it escapes the proton in the form of the final state interaction (\cite{Burkardt:2012sd}). The difference in $L^q$ due to different choices of  Wilson line configurations has been numerically confirmed in lattice QCD simulations (\cite{Engelhardt:2017miy,Engelhardt:2020qtg}).

Returning to (\ref{wigner}), it is tempting to define the parton distribution function for OAM  
\beq
L^{q,g}_{can}(x) = \int d^2b_\perp d^2k_\perp (b_\perp \times k_\perp)W_{q,g}(x,b_\perp,k_\perp) ,
\label{oam}
\eeq
where the staple-shaped Wilson line is used in $W_{q,g}$ to be consistent with the parton picture. One can show that (\ref{oam}) indeed leads to a satisfactory definition of PDF which generalizes the  phenomenologically introduced ones in the past. Using the equations of motion and the Lorentz invariant relations, one finds (\cite{Hatta:2012cs}, see also \cite{Ji:2012ba}))
\beq
L^q_{can}(x)&=&x\int_x^{\epsilon(x)}\frac{dx'}{x'}(q(x')+E_q(x'))-x\int_x^{\epsilon(x)}\frac{dx'}{x'^2}\Delta q(x')  \nn
&& -x\int_x^{\epsilon(x)}dx_1\int_{-1}^1 dx_2\Phi_F(x_1,x_2)P\frac{3x_1-x_2}{x_1^2(x_1-x_2)^2} - x\int_x^{\epsilon(x)}dx_1\int_{-1}^1 dx_2\tilde{\Phi}_F(x_1,x_2)P\frac{1}{x_1^2(x_1-x_2)} ,\label{t3}
\eeq
where $E_q$ is the E-type GPD defined in (\ref{gpd}) and $\epsilon(x)$ is the sign function. ($x$ can be negative, representing the antiquark OAM.) A similar formula for the gluon OAM distribution $L_{can}^g(x)$ is also available. The first line of  (\ref{t3}) is the Wandzura-Wilczek part related to twist-two distributions. In the second line, $\Phi_F$ and $\tilde{\Phi}_F$ are the `genuine' twist-three distributions defined by the quark-gluon-quark correlation function 
\beq
\int \frac{dy^-dz^-}{(2\pi)^2}e^{iP^+y^-\frac{x_1+x_2}{2}+iP^+z^-(x_2-x_1)}\langle P'S'|\bar{\psi}(-y^-/2)\gamma^+U gF^{+i}(z^-)U\psi(y^-/2)|PS\rangle = \frac{S^+}{P^+}\epsilon^{ij}\Delta_j \Phi_F(x_1,x_2)+\cdots, 
\eeq
and similarly for $\tilde{\Phi}_F$. 
This clearly shows that OAM is essentially a  twist-three observable. (\ref{t3}) also dictates how $L_{can}^{q,g}(x)$ evolve with the renormalization scale. Earlier works focused on the evolution of the Wandzura Wilczek part  
(\cite{Ji:1995cu,Hagler:1998kg,Harindranath:1998ve,Hatta:2018itc}). The evolution of the genuine twist-three part has been studied in \cite{Hatta:2019csj}.

\section{Helicity and orbital angular momentum at small-$x$}

At present, the largest uncertainties in the determination of $\Delta \Sigma$ and especially $\Delta G$ come from the small-$x$ region, say $x<10^{-2}$. While no experiments can probe the $x\to 0$ region, an accurate theoretical control over the behavior of the polarized parton distributions $\Delta q(x), \Delta G(x)$ is crucial for a better extraction of $\Delta G(x)$ at the future EIC. Mainly motivated by this situation, there has been renewed interest in the resummation of high energy logarithms $\ln\frac{1}{x}$ that is needed to describe the small-$x$ behavior of polarized PDFs (\cite{Kovchegov:2015pbl,Kovchegov:2017lsr,Kovchegov:2016weo,Kovchegov:2019rrz,Boussarie:2019icw,Kovchegov:2021lvz,Cougoulic:2022gbk,Adamiak:2023yhz,Adamiak:2023okq}). It is known that the nature of small-$x$ resummation is very different in the unpolarized and polarized distributions. This is already clear from inspection of the one-loop $q\to g$ splitting function
\beq
P_{gq}(x)=C_F\left(\frac{1}{x}+\frac{(1-x)^2}{x}\right), \qquad \Delta P_{gq}(x)=C_F\left(\frac{1}{x}-\frac{(1-x)^2}{x}\right).
\eeq
The two terms correspond to emissions of gluons with opposite helicities. In the unpolarized case, the two terms add up, leading to the well-known $1/x$ singularity in soft gluon emissions. However, this cancels in the polarized splitting function. Already from this fact, one can infer that  
\beq
G(x) \sim \frac{1}{x^{1+a}}, \qquad \Delta G(x) \sim \frac{1}{x^{b}}. \qquad (0< a,b<1) \label{det}
\eeq
Namely, the growth of $\Delta G(x)$ is much weaker than that of $G(x)$. However,  the value of $b$ still matters in the integral $\Delta G=\int_0^1 dx \Delta G(x)$.  In the unpolarized case, the value $a$ can be  determined from the Balitsky-Fadin-Kuraev-Lipatov (BFKL) equation (\cite{Kuraev:1977fs,Balitsky:1978ic}) that resums single logarithmic terms $(\alpha_s\ln \frac{1}{x})^n$. This is the famous BFKL Pomeron intercept 
\beq
a=4 \frac{\alpha_sN_c}{\pi}\ln 2,
\label{bfkl} \eeq
In the unpolarized case, the relevant resummation is double logarithmic $(\alpha_s\ln^2 \frac{1}{x})^n$. The second logarithm in $\frac{1}{x}$ comes from the transverse momentum integration $\int^{s}\frac{d^2k_\perp}{k_\perp^2}$.  Their resummation is conceptually more complicated and requires an ingenious theoretical tool called InfraRed Evolution Equation (IREE) (\cite{Kirschner:1983di}). The first determination of the parameter $b$ was achieved  in  \cite{Bartels:1995iu,Bartels:1996wc} 
\beq
b\approx 3.664 \sqrt{\frac{\alpha_sN_c}{2\pi}}
\eeq
The square-root is the hallmark of the double logarithmic resummation $e^{\sqrt{\alpha_s\ln^2\frac{1}{x}}}=\frac{1}{x^{\sqrt{\alpha_s}}}$. 
I quoted the number in the prefactor to the third decimal place because this has been recently challenged by   Borden and Kovchegov who instead found  (\cite{Borden:2023ugd,Borden:2024bxa,Borden:2025ehe}) 
\beq
b\approx 3.661\sqrt{\frac{\alpha_sN_c}{2\pi}}.
\eeq
This was obtained by analytically solving the small-$x$ evolution equation for the `polarized dipole amplitude' (\cite{Cougoulic:2022gbk}). 
The difference shows up at 4th order in perturbation theory, and is numerically very small. Presumably next-to-leading logarithmic  corrections, which are yet to be calculated  (see however, \cite{Ermolaev:2003zx,Kovchegov:2021lvz}), have a larger numerical impact in practice. But this subtlety, already appearing at leading logarithmic level, shows how nontrivial and challenging the resummation of helicity dependent PDFs is.

What about the small-$x$ behavior of the OAM PDFs $L_{can}^{q,g}(x)$?  Suppose an energetic quark emits a gluon carrying a small fraction $x\ll 1$ of its longitudinal momentum. Before and after the emission, the quark has essentially the same momentum and spin. Therefore the total angular momentum of the emitted gluon is zero, meaning that its helicity and OAM must cancel. This simple physical argument  survives after including  quantum effects (\cite{More:2017zqp,Hatta:2018itc}). To leading order, one finds a sign difference in the DGLAP evolution kernels
\beq
\frac{d}{d\ln Q^2} \Delta G(x) = \frac{\alpha_s}{2\pi} \int_x^1 \frac{dz}{z} (2C_F+\cdots )\Delta q(x/z), \qquad  
\frac{d}{d\ln Q^2} L^g_{can}(x) = \frac{\alpha_s}{2\pi} \int_x^1 \frac{dz}{z} (-2C_F+\cdots )\Delta q(x/z),
\eeq
which leads to the approximate relation $\Delta G(x)\approx -L^g_{can}(x)$ at small-$x$. 
There is also an all-order argument (\cite{Boussarie:2019icw}): If $\Delta G(x)$ behaves as in (\ref{det}), then 
\beq
L^g_{can}(x) \approx -\frac{2}{1+b}\Delta G(x) \qquad (x\to 0). \label{lgx}
\eeq
This has been derived from the gluonic version of (\ref{t3}) (\cite{Hatta:2012cs})
\beq
L^g_{can}(x)=\frac{x}{2}\int_x^{\epsilon(x)}\frac{dx'}{x'^2}(x'G(x')+E_g(x'))-x\int_x^{\epsilon(x)}\frac{dx'}{x'^2}\Delta G(x') +\cdots,
\eeq
where  only the Wandzura-Wilczek part is shown, assuming that the genuine twist-three part is subleading at small-$x$. Substituting  (\ref{det}) and further assuming that $a< b$ because $a \propto \alpha_s$ and $b\propto \sqrt{\alpha_s}$, one recovers (\ref{lgx}).  
A more careful analysis has been done in the polarized dipole amplitude framework  (\cite{Kovchegov:2023yzd,Manley:2024pcl}), again finding a large cancellation. Since the prefactor in (\ref{lgx}) is larger than unity, the minus sign means overcompensation. If, in future, the EIC discovers that the contribution to  $\Delta G$ from the small-$x$ region is  sizable, one should expect an even larger contribution to $L_{can}^g$ from the same $x$-region with an opposite sign. 

Finally, I briefly comment on the small-$x$ behavior of the GPDs $H_{q,g}(x,\xi)$ and $E_{q,g}(x,\xi)$ relevant to the Ji sum rule (\ref{jisum}), (\ref{jisum2}).  At zero skewness $\xi=0$, the small-$x$ behavior of $H_g(x,\xi=0)=xG(x)$ is governed by the BFKL exponent (\ref{bfkl}). It can be shown (\cite{Hatta:2022bxn}) that the same is true for the GPD $E_g(x,\xi=0)$. However, in practice, experiments probe GPDs at finite skewness $\xi\neq 0$, and there is an interesting interplay between $x$ and $\xi$ when both variables becomes small. Building on an earlier discussion in \cite{Bartels:1981jh}, a recent study   \cite{Kovchegov:2025yyl,Kovchegov:2026gwb} has concluded that the rapidity variable for the small-$x$ evolution of the quark and gluon GPDs $H_{q,g}(x,\xi),E_{q,g}(x,\xi)$ should be chosen as 
\beq
Y=\ln \left({\rm min} \left\{\frac{1}{|x|}, \frac{1}{\xi}\right\}\right). \qquad (|x|,\xi \ll 1)
\eeq
Thus the evolution in the DGLAP region $x>\xi$ is essentially unchanged, whereas  in the ERBL region $\xi>|x|$, there will be a power-law behavior in $1/\xi$.

\section{Observables for orbital angular momentum}

Finally, I come to a thorny question:  How can we experimentally measure the canonical OAM  $L_{can}^{q,g}$ of quarks and gluons necessary to complete the Jaffe-Manohar sum rule (\ref{jm})? For the extraction of the twist-two quantities $\Delta q$ and $\Delta G$,  collinear QCD factorization and global analyses provide a systematic framework which can be improved, in principle, to arbitrary precision.  Recently, next-to-next-to leading order (NNLO) global analyses have become available (\cite{Borsa:2024mss,Bertone:2024taw}). In contrast, for the OAMs, we are still at the stage of identifying observables that may be  sensitive to $L_{can}^{q,g}$. The first ideas were discussed about 10 years ago (\cite{Rajan:2016tlg,Ji:2016jgn,Hatta:2016aoc}),  but it is only recently that concrete numerical results for more promising observables became available (\cite{Bhattacharya:2022vvo,Bhattacharya:2023yvo,Bhattacharya:2023hbq,Bhattacharya:2024sck,Kovchegov:2024wjs,Bhattacharya:2026qnd}). The major difficulty in measuring OAMs is the fact that they are twist-three objects as explained above. Indeed, the definition $L_z=(b_\perp\times k_\perp)_z$ suggests that relevant observables must be sensitive to both  position $b_\perp$ and momentum $k_\perp$. Position $b_\perp$ is Fourier conjugate to the proton's recoil momentum $\Delta_\perp$ in exclusive processes. In addition, one has to measure the transverse momentum $P_\perp$ of  hadrons  or jets originating, at least in part, from the  intrinsic momentum $k_\perp$ of quarks and gluons. Observables that satisfy these requirements are longitudinal double spin  asymmetries in  coherent diffractive dijet production (\cite{Ji:2016jgn,Hatta:2016aoc,Bhattacharya:2022vvo,Bhattacharya:2024sck,Kovchegov:2024wjs})  and  the exclusive production of certain mesons (\cite{Bhattacharya:2023yvo,Bhattacharya:2023hbq}). However, at the moment, all these calculations have been  done only at leading order. For twist-three observables, even the leading order computation is often quite complicated.    Moreover, the evolution of the Wigner distribution, or its Fourier transform, the generalized TMD (GTMD), is still largely unexplored  (\cite{Echevarria:2016mrc,Bertone:2025vgy}). As the measurement of OAM is one of the key goals of the EIC, more resources should be invested to make further progress in this direction.

\section{Conclusions}

Thirty years after the EMC experiment, the partonic spin structure of the nucleon is now much better understood. Nowadays, the word ``crisis" is used only in a historical context. The Jaffe–Manohar and Ji decompositions offer complementary but not fully equivalent perspectives on how quark and gluon spin and orbital angular momentum  build up the nucleon’s total spin. There have been significant developments towards better understanding these sum rules during the past two decades, both in theory, global analysis and lattice QCD. Yet, further theoretical and phenomenological studies--particularly concerning  the observability of  OAM--are still needed to fully capitalize on the upcoming opportunities provided by the Electron-Ion Collider.

\begin{ack}[Acknowledgments]

The author is supported by the U.S. Department
of Energy under Contract No. DE-SC0012704, by LDRD funds from Brookhaven Science Associates.

\end{ack}


\bibliographystyle{Harvard}
\bibliography{reference}

\end{document}